\documentclass[%
 reprint,
superscriptaddress,
prab,
]{revtex4-1}

\usepackage{graphicx}
\usepackage{dcolumn}
\usepackage{bm}
\usepackage{amsmath}
\usepackage{siunitx}
\usepackage[english]{babel}
\usepackage{amsfonts}

\begin{document}

\preprint{APS/123-QED}

\title{Guiding of high-intensity laser pulses in \SI{100}{mm}-long hydrodynamic optical-field-ionized plasma channels
}

\author{A. Picksley}%
\affiliation{John Adams Institute for Accelerator Science and Department of Physics,University of Oxford, Denys Wilkinson Building, Keble Road, Oxford OX1 3RH, United Kingdom}%
\author{A. Alejo}%
\affiliation{John Adams Institute for Accelerator Science and Department of Physics,University of Oxford, Denys Wilkinson Building, Keble Road, Oxford OX1 3RH, United Kingdom}%
\author{J. Cowley}%
\affiliation{John Adams Institute for Accelerator Science and Department of Physics,University of Oxford, Denys Wilkinson Building, Keble Road, Oxford OX1 3RH, United Kingdom}%
\author{N. Bourgeois}%
\affiliation{Central Laser Facility, STFC Rutherford Appleton Laboratory, Didcot OX11 0QX, United Kingdom}%
\author{L. Corner}%
\affiliation{Cockcroft Institute for Accelerator Science and Technology, School of Engineering, The Quadrangle, University of Liverpool, Brownlow Hill, Liverpool L69 3GH, United Kingdom}%
\author{L. Feder}%
\affiliation{Institute for Research in Electronics and Applied Physics, University of Maryland, College Park, Maryland 20742, USA}%
\author{J. Holloway}%
\affiliation{John Adams Institute for Accelerator Science and Department of Physics,University of Oxford, Denys Wilkinson Building, Keble Road, Oxford OX1 3RH, United Kingdom}%
\author{H. Jones}%
\affiliation{Cockcroft Institute for Accelerator Science and Technology, School of Engineering, The Quadrangle, University of Liverpool, Brownlow Hill, Liverpool L69 3GH, United Kingdom}%
\author{J. Jonnerby}%
\affiliation{John Adams Institute for Accelerator Science and Department of Physics,University of Oxford, Denys Wilkinson Building, Keble Road, Oxford OX1 3RH, United Kingdom}%
\author{H. M. Milchberg}%
\affiliation{Institute for Research in Electronics and Applied Physics, University of Maryland, College Park, Maryland 20742, USA}%
\author{L. R. Reid}%
\affiliation{Cockcroft Institute for Accelerator Science and Technology, School of Engineering, The Quadrangle, University of Liverpool, Brownlow Hill, Liverpool L69 3GH, United Kingdom}%
\author{A. J. Ross}%
\affiliation{John Adams Institute for Accelerator Science and Department of Physics,University of Oxford, Denys Wilkinson Building, Keble Road, Oxford OX1 3RH, United Kingdom}%
\author{R. Walczak}%
\affiliation{John Adams Institute for Accelerator Science and Department of Physics,University of Oxford, Denys Wilkinson Building, Keble Road, Oxford OX1 3RH, United Kingdom}%
\author{S. M. Hooker} \email{simon.hooker@physics.ox.ac.uk}
\affiliation{John Adams Institute for Accelerator Science and Department of Physics,University of Oxford, Denys Wilkinson Building, Keble Road, Oxford OX1 3RH, United Kingdom}%

\date{\today}

\begin{abstract}
Hydrodynamic optically-field-ionized (HOFI) plasma channels up to \SI{100}{mm} long are investigated. Optical guiding is demonstrated of laser pulses with a peak input intensity of $\SI{6E17}{W.cm^{-2}}$ through \SI{100}{mm} long plasma channels with on-axis densities measured interferometrically to be as low as $n_{e0} = \SI[separate-uncertainty=true]{1.0(3)e17}{cm^{-3}}$. Guiding is also observed at lower axial densities, which are inferred from magneto-hydrodynamic simulations to be approximately  $\SI{7e16}{cm^{-3}}$. Measurements of the power attenuation lengths of the channels are shown to be in good agreement with those calculated from the measured transverse electron density profiles. To our knowledge, the plasma channels investigated in this work are the longest, and have the lowest on-axis density, of any free-standing waveguide demonstrated to guide laser pulses with intensities above $>\SI{E17}{W.cm^{-2}}$.

This article was published in Physical Review Accelerators and Beams \textbf{23}, 081303 on 26 August
2020. DOI: 10.1103/PhysRevAccelBeams.23.081303

\textcopyright  2020 American Physical Society.
\end{abstract}

\maketitle

\section{Introduction}\label{sec:intro}
Since they can generate acceleration gradients three orders of magnitude greater than conventional radio-frequency machines, plasma accelerators offer a potential route to a new generation of compact radiation sources and, in the longer term, compact particle colliders.

Laser-driven \cite{Leemans2006, Kneip:2009, Wang:2013el, Leemans2014, Gonsalves:2019ht} and particle-driven plasma accelerators \cite{Blumenfeld2007, Litos2014, adli2018} can now routinely generate bunches of electrons with GeV-scale energies, which has stimulated considerable interest in constructing the first generation of facilities driven by plasma accelerators. For example, the EuPRAXIA project \cite{Walker:2017hi} envisages laser- and particle-driven plasma accelerators delivering \SI{5}{GeV} electron bunches, at a repetition rate above \SI{5}{Hz}, with sufficient quality to drive a free-electron laser operating at wavelengths below \SI{36}{nm}. Similar operating parameters have been identified in the US roadmap for advanced accelerators \cite{USroadmap}.

\begin{figure*}[t] 
    \centering
    \includegraphics[width=170mm]{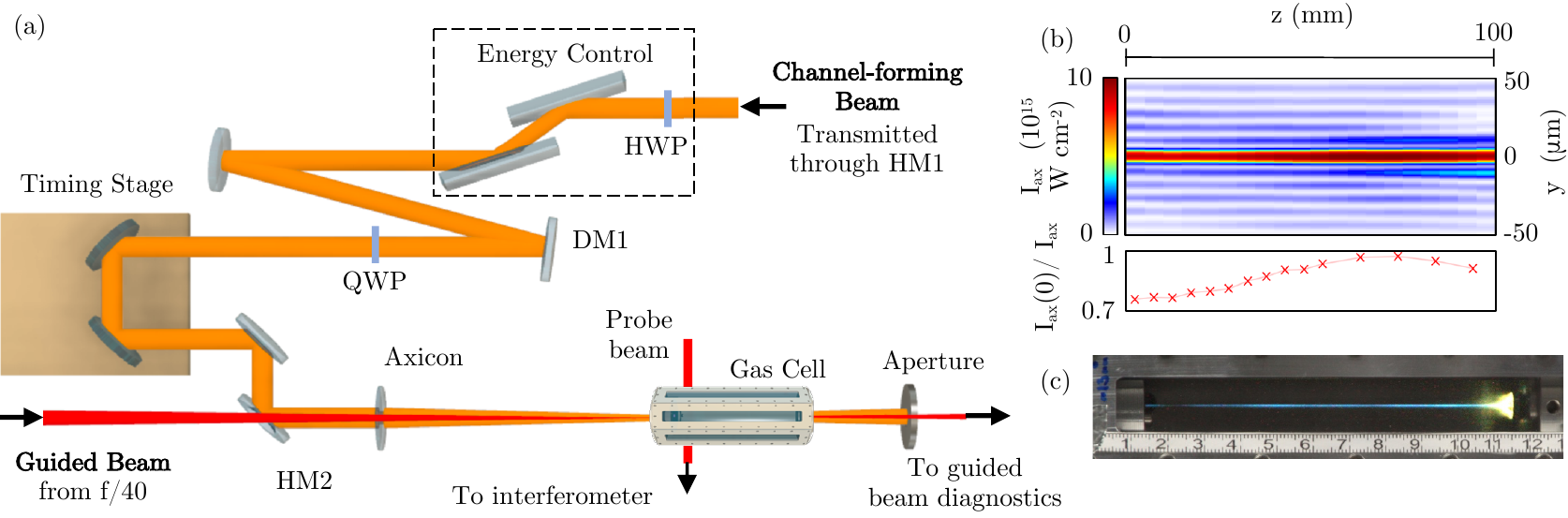}
    \caption{[color online] (a) Schematic diagram of the experiment layout. (b) Longitudinal variation of the transverse intensity profile of the axicon focus, measured in vacuo by a camera in the vacuum chamber. The red curve shows the axial intensity $I_{ax}(0)$ as a function of longitudinal position. (c) Time-integrated image of the visible plasma emission produced by the channel-forming beam focused into the gas cell at a fill pressure $P = \SI{26}{mbar}$. The scale visible at the bottom of the image is in cm. Note that the apparent decrease in plasma brightness near a scale reading of \SI{2.5}{cm} arises from blackening of the cell window in that region, not from non-uniformity of the plasma.}
    \label{fig:exp_setup}
\end{figure*}    

Design studies for multi-GeV laser-driven plasma accelerator facilities propose operation in the quasi-linear regime, in plasma stages of density $n_e \approx \SI{1E17}{cm^{-3}}$ and a length in the range \SIrange{0.25}{1.}{m}. This length is many times the Rayleigh range $z_R$ of the drive pulse, and, since self-guiding does not occur in this regime, the drive laser pulse must be guided in a waveguide structure.

The development of waveguides with properties suitable for plasma accelerators has been an active area of research for several decades. Guiding of high-intensity laser pulses has been demonstrated in hollow capillaries \cite{Cros:2002, Wojda:2009, Genoud:2011kj} and in plasma channels. A wide variety of methods for generating plasma channels has been investigated, including: hydrodynamic expansion of laser-heated plasma columns \cite{Durfee:1993, Durfee:1994wz, Durfee:1995gr, Volfbeyn:1999cj}; fast discharges in capillaries \cite{Hosokai:2000, Luther:2005eu,Wang:2005ea} or in an open geometry \cite{Lopes:2003}; and  slow discharges in ablated \cite{Ehrlich:1996, Kaganovich:1999} or gas-filled capillaries \cite{Spence:2000fr, Butler:2002zza, Gonsalves:2007}. For applications to laser-driven plasma accelerators, the most successful approach has been the gas-filled capillary discharge waveguide \cite{Spence:2000fr, Butler:2002zza, Gonsalves:2007}. This was used in the first demonstration \cite{Leemans2006} of laser-driven acceleration to \SI{1}{GeV}, and in later work \cite{Leemans2014} showing acceleration to \SI{4.2}{GeV}. Capillary discharge waveguides have been operated at a pulse repetition rate as high as \SI{1}{kHz} \cite{Gonsalves:2016jc}. The use of an additional laser heater to deepen the plasma channel formed in low-density capillary discharge waveguides has been investigated theoretically \cite{Bobrova:2013jz}. Recently this method was used to generate plasma channels with an on-axis density of $n_{e0} \approx \SI{3E17}{cm^{-3}}$ and a matched spot size $W_M \approx \SI{60}{\micro \metre}$; these channels were used to accelerate electrons to \SI{7.8}{GeV} \cite{Gonsalves:2019ht}.  However, despite this considerable success, it is not clear whether capillary discharge waveguides would be suitable for future plasma accelerators operating at high pulse repetition rates, when laser damage to the capillary and discharge structure is likely to be an issue.

Hydrodynamic plasma channels \cite{Durfee1993, Durfee:1995gr, Sheng2005} offer an alternative approach. Since these are free-standing, with no close-lying physical structure which could be damaged by the guided laser pulse, they are well suited to long-term operation at high pulse repetition rates. In a hydrodynamic channel a column of plasma is formed, and collisionally-heated to a few tens of eV \cite{Clark1998}, by a picosecond-duration laser pulse focused with an axicon lens. The hot plasma column expands rapidly, driving a shock wave into the surrounding cold gas, forming a plasma channel within the cylindrical shock front.  A remaining issue is that rapid collisional heating of the initial plasma column requires a high gas density, which results in a relatively high on-axis plasma channel density of order $\SI{1E18}{cm^{-3}}$; this is too high for multi-GeV accelerator stages. 

To overcome this limitation we have proposed \cite{Shalloo:2018fy} a variation of this approach --- hydrodynamic optical-field-ionized (HOFI) plasma channels --- in which the initial plasma column is formed by optical field ionization.  The energies of the electrons produced by optical field ionization (OFI) are independent of the gas density, and can be controlled by varying the polarization of the ionizing laser. This allows the generation of plasma channels of significantly lower density. We note that Lemos et al.\ have also investigated hydrodynamic channels generated by field-ionized plasmas \cite{Lemos:2013gb, Lemos:2013ju, Lemos:2018gx}, and recently Smartsev et al.\ used an axiparabola to generate \SI{10}{mm} long channels of this type \cite{Smartsev:2019cy}.

Previously we demonstrated \cite{Shalloo:2019hv} the formation of, and guiding in, \SI{16}{mm} long HOFI channels produced by an axicon lens, with an on-axis density as low as $n_e(0) = \SI{1.5E17}{cm^{-3}}$.  Control of the channel parameters by variation of the initial gas pressure, and the delay after the arrival of the channel-forming pulse, was demonstrated. High-quality guiding was achieved at a pulse repetition rate of \SI{5}{Hz}, limited by the available channel-forming laser and vacuum pumping system.

In this work, we extend HOFI plasma channels to a length of \SI{100}{mm} for the first time. We demonstrate optical guiding of laser pulses with a peak input intensity of $\SI{6E17}{W.cm^{-2}}$ through \SI{100}{mm} long plasma channels with interferometrically-measured on-axis densities as low as $n_{e0} = \SI[separate-uncertainty=true]{1.0(3)e17}{cm^{-3}}$. We also present results which demonstrate guiding at lower axial densities, which are estimated to be below $\SI{1e17}{cm^{-3}}$.  Measurements of the power attenuation lengths of these channels are in good agreement with those calculated from the measured transverse electron density profiles. To our knowledge, the plasma channels investigated in this work are the longest, and have the lowest on-axis density, of any free-standing waveguide demonstrated to guide laser pulses with an intensity above $\SI{E17}{W.cm^{-2}}$.

\section{Experimental Setup}\label{sec:setup}
The experiment was performed with the Astra-Gemini TA3 Ti:sapphire laser at the Rutherford Appleton Laboratory, UK. This laser system provides two, linearly-polarized beams, each delivering optical pulses with a central wavelength of $800 \ \mathrm{nm}$ and a full-width at half-maximum  duration of \SI[separate-uncertainty = true]{40(3)}{fs}.

Figure \ref{fig:exp_setup}(a) shows schematically the experiment setup employed (further details are given in the Supplemental Information). Both the channel-forming beam and the beam to be guided were formed from a single Astra-Gemini beam. The guided beam was formed by reflection from a high-reflectivity dielectric mirror (HM1, not shown in Fig.\ \ref{fig:exp_setup}) in which a hole had been drilled. The pulse energy in this beam was then reduced by reflection from an uncoated optical wedge to approximately \SI{1.1}{J}. The beam transmitted by HM1, constituted the channel-forming beam, and comprised laser pulses with an energy of up to \SI{160}{mJ}. 

The channel-forming beam was passed through a pulse energy control system, its wavefront flattened by reflection from a deformable mirror (DM1), and its polarization converted to circular \cite{Shalloo:2018fy,Shalloo:2019hv}.  It was then sent to a retro-reflecting timing stage, reflected by a second holed mirror (HM2), and focused into the target gas cell by a fused silica axicon lens of base angle $\vartheta = \SI{5.6}{\degree}$. The cell was filled with hydrogen gas to an initial pressure $P$ which could be varied in the range \SIrange{10}{70}{mbar}.

The wavefront of the guided beam was corrected by a second deformable mirror (DM2), and directed to an $f = \SI{6}{m}$ off-axis paraboloid, used at $f/40$, which focused the beam into the gas cell. Separate measurements  of the transverse fluence profile as a function of longitudinal position showed that in vacuo the guided beam had a waist of  $w_0 = \SI{38}{\mu m}$ (radius of $1/\mathrm{e}^2$ relative intensity) and a Rayleigh range of $z_\mathrm{R} = \SI[separate-uncertainty=true]{4.5(8)}{mm}$.

Both channel-forming and guided beams entered and left the gas cell via pinholes located at each end of the cell. The length of an axicon focus is proportional to the diameter of the limiting aperture, which is usually that of the axicon or the input beam. In our case the limiting aperture was the entrance pinhole, and hence the diameter of this pinhole determined the maximum length of the channel. Two different entrance pinhole diameters were used: \SI{5}{mm} and \SI{10}{mm}, limiting the channel lengths to \SI{50}{mm} and \SI{100}{mm} respectively. When the \SI{10}{mm} pinhole was used, the increased outflow of gas limited the fill pressure to a maximum value of \SI{26}{mbar}. The diameter of the exit pinhole was \SI{1}{mm} for all the work reported here.

The longitudinal position of the gas cell was adjusted to achieve the best coupling of the guided beam. The optimal focal position of the guided beam was found to be within one $z_{\mathrm{R}}$ of the entrance pinhole. On leaving the gas cell the guided beam was reduced in intensity by reflections from an uncoated optical wedge, and imaged by a reflective Keplerian telescope onto a 16-bit CCD camera.

Figure \ref{fig:exp_setup}(b) shows, as a function of the longitudinal distance $z$ from the entrance pinhole, the  transverse fluence of the channel-forming beam, measured with the cell removed. It can be seen that the transverse intensity profile $I_\mathrm{ax}(r)$ of the channel-forming beam was essentially independent of $z$ over a distance of \SI{100}{mm}. Analysis of this profile shows that it was close to that expected for an axicon illuminated by a top-hat beam \cite{McLeod:1954}: $I_\mathrm{ax}(r) \propto J_0^2(\beta r)$, where $\beta = k[\arcsin(\eta \sin \vartheta) - \vartheta]$, $k = 2\pi / \lambda$ is the wavenumber of the incident light, and $\eta$ is the refractive index of the axicon substrate. The first intensity minimum occurs at a radius $r = \SI{7.1}{\micro \metre}$, in close agreement with that expected from the calculated value of $\beta$. Consistent with this \cite{Herman:91}, the peak intensity of the focused channel-forming beam increased with longitudinal position, from $\SI{7E15}{W.cm^{-2}}$, between  $z = 0$ and $z = \SI{100}{mm}$. Simulations showed that dispersion in the material of the axicon stretched the pulse by approximately \SI{40}{fs} and therefore reduced the peak intensity by a factor of approximately two \cite{ross2020}. This decrease in intensity is not expected to change the properties of the focus or plasma formation substantially.  Figure \ref{fig:exp_setup}(c) shows a time-integrated image of the visible plasma emission, for $P = \SI{26}{mbar}$, when only the channel-forming beam entered the cell. It can be seen clearly that the brightness and diameter of the plasma were nearly uniform over the \SI{100}{mm} length of the cell; the uniformity of the initial plasma is a consequence of the strong dependence on intensity of optical field ionization, and the fact that the transverse intensity profile and the axial intensity of the channel-forming beam depend only weakly on $z$.

The transverse electron density profile of the plasma was measured by folded-wave interferometry, using \SI{800}{nm}, \SI{40}{fs} probe pulses derived from the second Astra-Gemini beam. The delay $\tau$ between the arrival at the gas cell of the channel-forming and probe beams could be adjusted in the range $\tau = $ \SIrange{0}{5}{ns} by a retro-reflecting timing stage in the probe beam line. Owing to the very small phase shift imparted on the probe beam ($< \SI{90}{mrad}$), it was not possible to measure the on-axis plasma density of channels formed at cell pressures below approximately \SI{25}{mbar}.

\section{Results}

\begin{figure}[b]
    \centering
    \includegraphics[width=84mm]{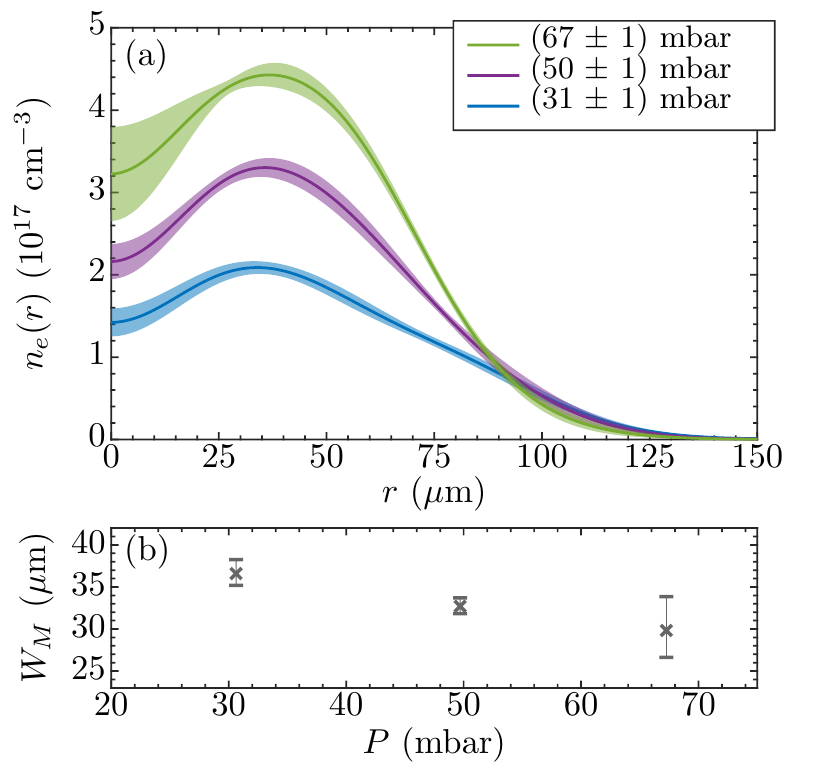}
    \caption{[color online] (a) Deduced electron density profiles  of HOFI channels for several initial fill pressures and $\tau = \SI{3.0}{ns}$. For each plot the light bands show the standard deviation obtained from averaging approximately 10 shots. (b) Variation of the measured matched spot size $W_M$ with fill pressure.}
    \label{fig:pressure_scan}
\end{figure}

\begin{figure*}[tb]
    \centering
    \includegraphics[width=170mm]{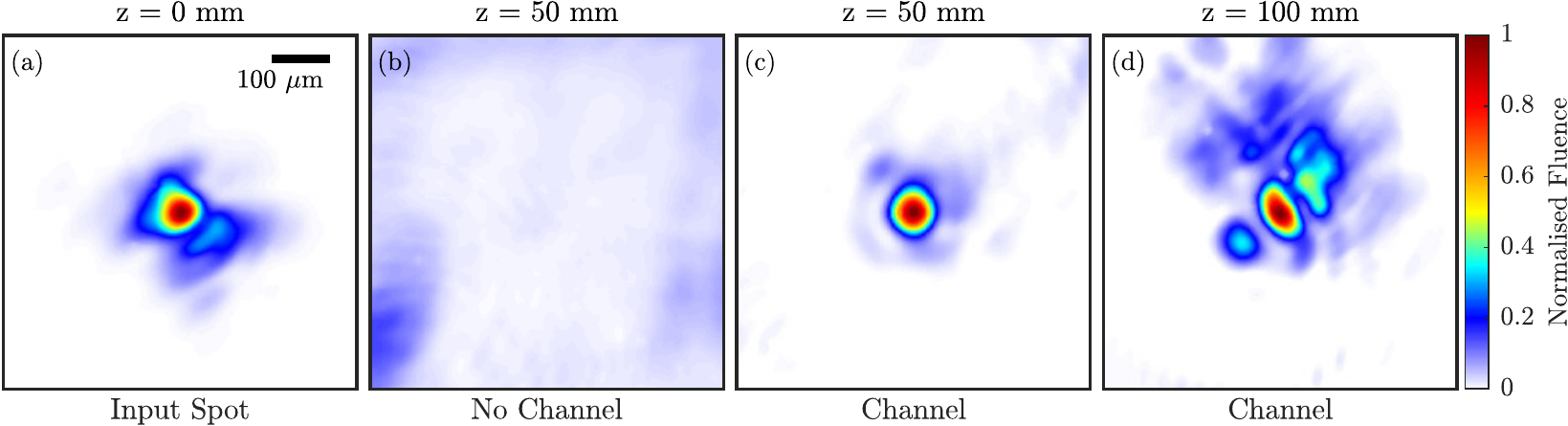}
    \caption{[color online] Measured transverse fluence profiles of the guided beam at: (a) focus, in vacuum; (b) $z = \SI{50}{mm}$, in vacuum; (c) $z = \SI{50}{mm}$, for $P = \SI{68}{mbar}$ and $\tau = \SI{3.0}{ns}$; (d) $z = \SI{100}{mm}$, for $P = \SI{26}{mbar}$ and $\tau = \SI{2.7}{ns}$. The transverse scale is the same for all plots, as indicated by the scale bar shown in (a). For plots (a), (c), and (d) the fluence is normalized to the peak value in that plot; the fluence scale for (b) is the same as in (a). Compared to (a), the fluence scales of (c) and (d) were increased by factors of approximately 4 and 7 respectively. }
    \label{fig:guiding_demo}
\end{figure*}

The transverse electron density profiles of the HOFI plasma channels were deduced from transverse interferometry using the  method described previously  \cite{Shalloo:2019hv} and outlined in the Supplemental Information. The temporal evolution of this profile was found to be consistent with our earlier work \cite{Shalloo:2018fy, Shalloo:2019hv}: at $\tau \approx 0$ a cylindrical column of plasma, with a diameter of approximately \SI{15}{\micro \metre} was formed; this  expanded rapidly, driving a cylindrical shock wave into the surrounding gas, and forming a plasma channel within the region encircled by the shock front. The plasma density profile evolved into channels suitable for guiding the input spot at times in the range $\SI{2.5}{ns} \lesssim \tau \lesssim \SI{3.5}{ns}$. Figure \ref{fig:pressure_scan} shows the measured transverse electron density profiles at $\tau = \SI{3.0}{ns}$ for several initial fill pressures, averaged over a longitudinal distance $\Delta z = \SI{1.2}{mm}$ centered at $z \approx \SI{25}{mm}$. It can be seen that the on-axis electron density $n_{e0}$ increases approximately linearly with $P$, as observed previously \cite{Shalloo:2019hv}, and that the position of the shock front is approximately independent of $P$, as expected from Sedov-Taylor blast-wave theory \cite{Hutchens2000}.  For $P = \SI{67}{mbar}$, the channel depth was measured to be $\Delta n_e = n_e(r_\mathrm{shock}) - n_{e0} = \SI[separate-uncertainty=true]{1.2(2)e17}{cm^{-3}}$ where $r_\mathrm{shock}$ is the measured radial position of the shock front. The channel depth is reduced to $\Delta n_e = \SI[separate-uncertainty=true]{0.7(1)e17}{cm^{-3}}$ for an initial fill pressure of $P = \SI{31}{mbar}$. For a finite channel comprising a parabolic electron density profile for radii $r < r_m$, and a constant density for $r > r_m$, only modes satisfying the following relation are expected to be guided with low loss \cite{Durfee:1994wz, Durfee:1995gr}: $(2p + m +1)^2 < \pi r_e r_m^2 \Delta n_e$, where $p$ and $m$ are the radial and azimuthal indices and $r_e$ is the classical electron radius. Taking $r_m$ to be equal to $r_\mathrm{shock}$, we find $(2p + m +1) \lesssim 1.3$ and $(2p + m +1) \lesssim 1.0$ for the $P=\SI{67}{mbar}$ and $P=\SI{31}{mbar}$ channels  respectively. It is therefore expected that only low-order modes will propagate without significant loss. The matched spot size of the lowest-order mode is given by $W_M^4 \approx r_\mathrm{shock}^2/\pi r_e \Delta n_e$; hence for these channels $W_M \approx \SI{30}{\micro\metre}$.

Figure \ref{fig:guiding_demo} demonstrates guiding of high-intensity laser pulses in HOFI channels up to \SI{100}{mm} long. Figure \ref{fig:guiding_demo}(a) shows the transverse fluence profile of the guided beam at focus, recorded by the forward diagnostic camera at full power. It can be seen that, despite the use of a deformable mirror, the input beam exhibited significant transverse structure. From this measured fluence profile the peak focused intensity is estimated to be \SI{6.1e17}{W.cm^{-2}}.

Guiding was investigated in \SI{50}{mm} and \SI{100}{mm} long channels. For the shorter channels the guiding was approximately optimized by adjusting the initial cell pressure $P$ and the delay $\tau$. For the input beam used in this work, the best guiding was found to occur at $P\approx \SI{68}{mbar}$ and $\tau \approx \SI{3.0}{ns}$. For the \SI{100}{mm} long channels the cell pressure was limited to $P \leq \SI{26}{mbar}$; at this lower pressure the optimum delay was found to be $\tau \approx \SI{2.7}{ns}$.

Figure \ref{fig:guiding_demo}(b) shows the profile of the beam at $z = \SI{50}{mm}$ in the absence of a HOFI channel, showing clearly the effect of diffraction over a distance of more than $10 z_\mathrm{R}$. Figure \ref{fig:guiding_demo}(c) shows the fluence profile in the same plane, but at a delay $\tau = \SI{3.0}{ns}$ after focusing the channel-forming beam into the cell filled to $P = \SI{68}{mbar}$. The increase in the peak transmitted fluence is striking, and demonstrates clearly that the pulse was guided through the  plasma channel generated by the channel-forming beam. For this pressure and timing, the on-axis density was measured to be $n_{e0} = \SI[separate-uncertainty=true]{3.8(5)e17}{cm^{-3}}$ and matched spot size $W_M = \left(30\,_{-3}^{+5}\right) \SI{}{\micro m}$ (see figure \ref{fig:pressure_scan}). The energy transmission was measured to be \SI{27}{\%} using the method described in the Supplemental Information.

Although a blocking aperture was used, some light from the channel-forming beam reached the mode-imaging camera. The signal from the channel-forming beam, and other contributions to the background, were removed as described in the Supplemental Information. However, this procedure was not perfect, and hence regions in which the signal is below 5-10 \% of the peak signal are likely to be contaminated by contributions from the channel-forming beam.

Fig.\ \ref{fig:guiding_demo}(d) shows the transverse fluence profile of the guided beam at the exit of a \SI{100}{mm} long HOFI channel formed at a similar delay ($\tau = \SI{2.7}{ns}$), but at a lower cell pressure of \SI{26}{mbar}. For this shot, the measured on-axis electron density was $n_{e0} = \SI[separate-uncertainty=true]{1.0(3)e17}{cm^{-3}}$ and matched spot size $W_M = \SI[separate-uncertainty=true]{34(7)}{\micro m}$. It is clear that guiding was also achieved in this case, the radius of the guided pulse was measured to be $w_\mathrm{out} = \SI[separate-uncertainty=true]{48(14)}{\micro m}$, close to that of the focal spot. The measured energy transmission was 14 \%. The low intensity light in the transverse wings can be attributed to the axicon light that reached the CCD, which is relatively more intense than in Fig.\ \ref{fig:guiding_demo}(c), or scattering from partially ionized gas at the exit of the waveguide.

\begin{figure}[tbh]
    \centering
    \includegraphics[width=84mm]{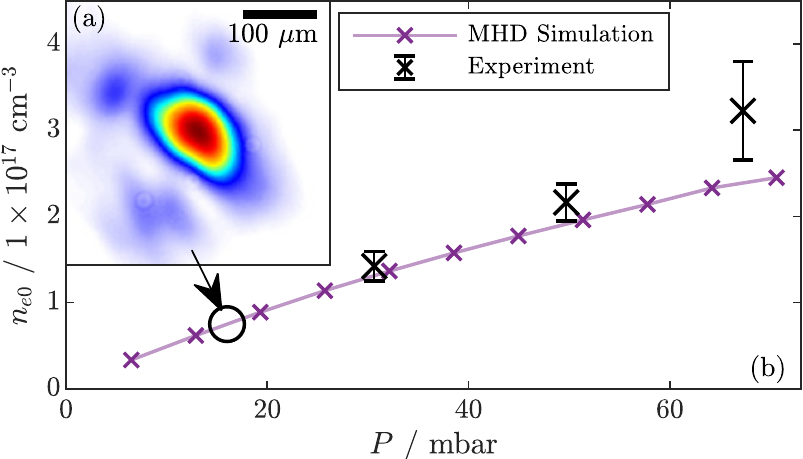}
    \caption{[color online] (a) Measured transverse fluence profile of the guided beam at the exit of  a \SI{50}{mm} long HOFI plasma channel formed at $\tau = \SI{3.0}{ns}$ and $P = \SI[separate-uncertainty=true]{17(1)}{mbar}$. (b) Comparison of the on-axis electron density measured by interferometry (symbols) and that calculated by MHD simulations (solid line) for $\tau = \SI{3.0}{ns}$ and $P = \SI{17}{mbar}$. The axial electron density calculated by the MHD simulations for $P = \SI{17}{mbar}$ is $\SI{7e16}{cm^{-3}}$.}
    \label{fig:lowest_guided}
\end{figure}

Optical guiding was also observed in 50 mm long channels  for cell pressures as low as \SI{17}{mbar}, as demonstrated in Fig.\ \ref{fig:lowest_guided}(a), which shows high-quality guiding, with a measured energy transmission of 29 \%. It was not possible to measure the electron density profile interferometrically, however, the on-axis density can be estimated with the magneto-hydrodynamic (MHD) code HELIOS \cite{MacFarlane2006}, assuming cylindrical symmetry. Figure \ref{fig:lowest_guided}(b) compares the on-axis electron density at $\tau = \SI{3.0}{ns}$ measured interferometrically with that calculated by the MHD simulations. It can be seen that the values of $n_{e0}$ measured for $P > \SI{26}{mbar}$ are in good agreement with the simulations. From the MHD simulations the on-axis electron density for the conditions of the guided spot shown in (a) is estimated to be $n_{e0} = \SI{7e16}{cm^{-3}}$.

Figure \ref{fig:length_scan} shows the energy transmission, $T(z)$, measured for several channel lengths and $P=\SI{26}{mbar}$ and $\tau = \SI{2.7}{ns}$. A total of 15 laser shots were recorded for each channel length, and those with an input pointing outside the measured acceptance angle of the channel ($\approx \SI{3.7}{mrad}$) were discarded. It can be seen that the experimental data show an approximately exponential decrease in $T(z)$ with $z$; a fit of the expression $T(z) = T(0)\exp\left( -z / L_\mathrm{att}\right)$ to the data yields a coupling efficiency  $T(0) = \left(45\,_{-11}^{+14}\right)\%$ and $L_\mathrm{att} = \left(102\pm38\right)\SI{}{mm}$.

Solving the paraxial Helmholtz equation \cite{Fan:2000fh, Clark:2000dk} numerically allows the measured variation of the transmission with $z$ to be compared with that expected for the measured transverse electron density profile of the channel (see Supplemental Information). Figure \ref{fig:length_scan} shows the results of these simulations for a channel with the transverse electron density profile measured for $P=\SI{26}{mbar}$ and $\tau = \SI{2.7}{ns}$.  For a Gaussian input beam, with spot size matched to that of the lowest-order mode of the channel, the coupling efficiency and attenuation length are found to be $T(0) \approx 100\%$ and  $L_{\mathrm{att}} = \SI[separate-uncertainty=true]{84(2)}{mm}$. This attenuation length is close to that observed in the experiment, but the coupling efficiency is much higher than is measured. This suggests that the lower measured values of $T(z)$ arise from the non-ideal transverse profile of the guided beam. To confirm this we used the same propagation code to simulate the propagation of a beam with an input transverse profile equal to that used in the experiment (see Fig.\ \ref{fig:guiding_demo}(a)), assuming a constant transverse spatial phase. As shown in Fig.\ \ref{fig:length_scan}, the calculated energy transmission for the real input beam agrees well with the measured data. We note that the non-exponential variation in $T(z)$ observed for $z \lesssim \SI{30}{mm}$ arises from the excitation of higher-order modes, which have higher propagation losses; as expected, for larger values of $z$ the rate of decrease of $T(z)$ closely follows that of the lowest-order mode. The attenuation length and coupling efficiency for the experimentally-measured input beam was deduced by fitting an exponential decay to the calculated transmission in the region $z\geq\SI{50}{mm}$, where the guided beam is dominated by the lowest order mode. The attenuation length is found to be $L_\mathrm{att} = \SI[separate-uncertainty=true]{84(2)}{mm}$, which agrees with the measured value to within errors, and is equal to that calculated for a matched input beam. The calculated coupling efficiency for the experimentally-measured input beam is $T(0) = \SI[separate-uncertainty=true]{48(4)}{\%}$. This coupling efficiency agrees well with an analysis of the input spot shown in Fig.\ \ref{fig:guiding_demo}(a) as an expansion of Laguerre-Gauss modes with a spot size equal to the matched spot size of the channel (see Supplemental Information): this shows the  fractional power contained in the lowest-order Laguerre-Gauss mode is $\SI[separate-uncertainty=true]{49(8)}{\%}$.

\section{Discussion}
We now make some further observations. First, the results presented in this work would not have been affected significantly by wakefield excitation or relativistic self-guiding. The guided pulse will have driven a plasma wave with a relative electron density of up to $\delta n_e/ n_e \sim 5 \%$, leading to a laser energy loss of approximately \SI{1}{mJ} per mm of plasma. Hence less than 9\% of the energy of the guided pulse will have been transferred to the plasma wave over a \SI{100}{mm} long channel. Second, relativistic self-focusing would not have been significant for the conditions of these experiments since, for the range of axial channel densities investigated, $0.1 \lesssim P_0 /P_\mathrm{crit} \lesssim 0.3$ where $P_0$ is the peak laser power and the critical power $P_\mathrm{crit} = 17.4(\omega_0/\omega_p)^2$, in which $\omega_0$ and $\omega_p$ are the laser and plasma frequencies respectively.

\begin{figure}[btp]
    \centering
    \includegraphics[width=84mm]{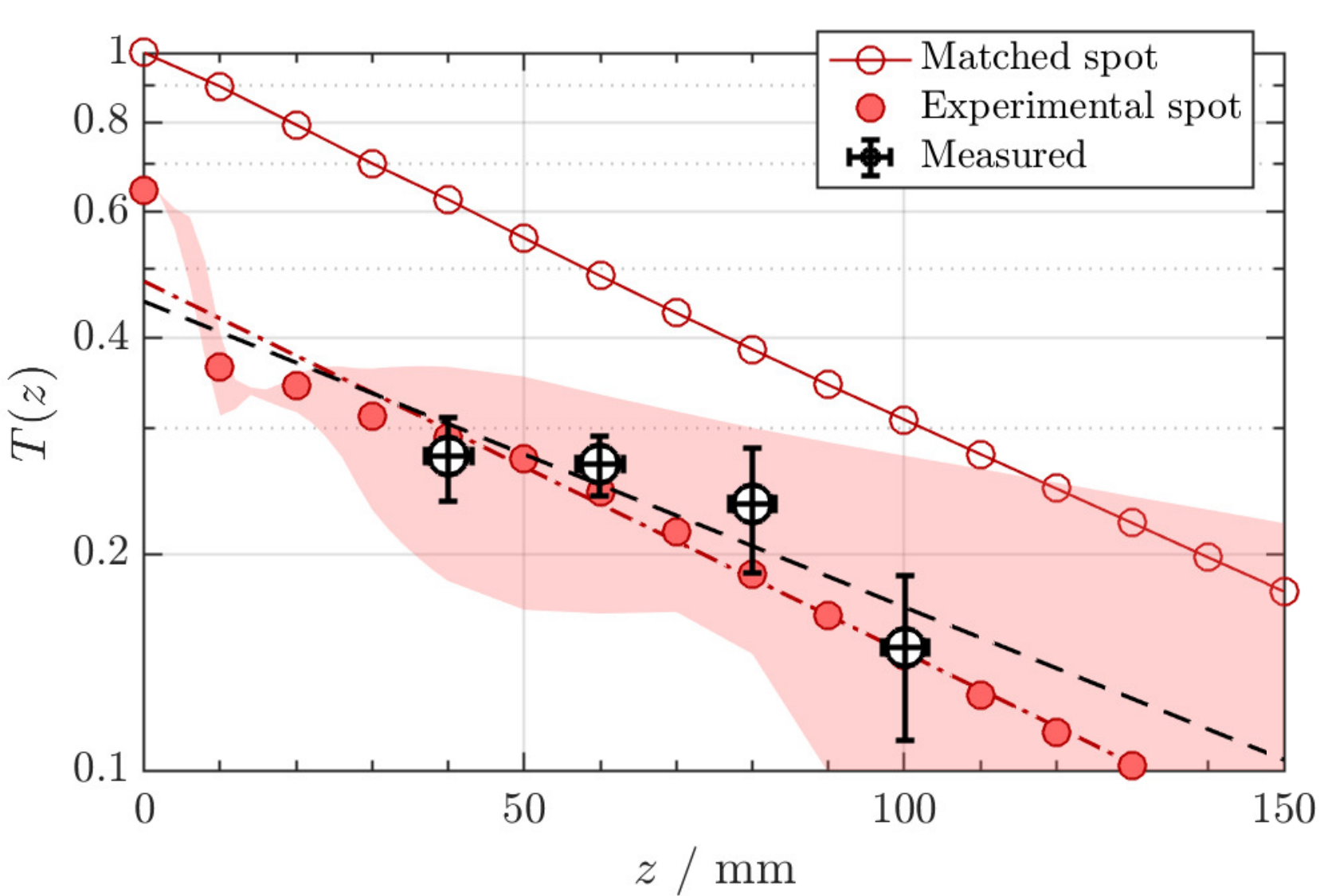}
    \caption{[color online] Comparison of the measured and simulated energy transmission for plasma channels formed at $P=\SI{26}{mbar}$ and $\tau = \SI{2.7}{ns}$. The measured data is shown as open black circles, with error bars with a total length equal to one standard deviation. The dashed black line shows a fit of the function $T_\mathrm{theory}(z) = T(0)\exp(-z / L_\mathrm{att})$ to the experimental data. The open red circles show the calculated transmission for a Gaussian input beams with a spot size matched to the lowest-order mode of the channel. The solid red circles show the calculated energy transmission for an input beam with a transverse intensity profile equal to the experimentally-measured profile (see Fig.\ \ref{fig:guiding_demo}(a)); the red shading shows the uncertainty in $T(z)$ arising from the the uncertainties in the electron density profile. The red dash-dot line shows a fit of the function $T_\mathrm{theory}(z)$ to the calculated transmission for the realistic input spot in the region $z > \SI{50}{mm}$.}
    \label{fig:length_scan}
\end{figure}

For this work, the channel-forming and guided beams were coupled into the gas cell via pinholes, and these needed to be large in order to create long channels. The plumes created by gas flow through the pinholes would have had a longitudinal scale length comparable to the pinhole diameter, which in turn $\approx 2z_{\mathrm{R}}$. As such the plume regions could in principle have played a role in coupling the guided beam into the main channel \cite{Dimitrov2007, Kim:2002}. The plasma channel would have extended into the entrance plume since the axicon focus extended into this region. For \emph{collisionally-heated} hydrodynamic channels the density-dependent heating rate causes the channel to narrow near its entrance \cite{Kim:2002}, which can decrease the coupling efficiency. However, narrowing of this type is unlikely to happen for HOFI channels since the OFI heating is independent of gas density; this is evident in Fig.\ \ref{fig:pressure_scan}, which shows that the radial position of the shock front at a given delay $\tau$ is approximately independent of density. The depth and axial density of the HOFI channel formed in the plume region are both expected to decrease with distance away from the cell. These properties will cause the matched spot to increase with distance from the cell, which may lead to improved coupling compared to collisionally-heated hydrodynamic channels. Studies of these end effects, and the development of improved cell designs, will be undertaken in future work.

It is possible that the guiding observed in this work may have been assisted by ionization of the neutral or partially-ionized gas surrounding the plasma channel, since ionization of this type would have increased the channel depth. The guiding simulations (see Supplemental Information) do not include this effect, but they do show that the laser field immediately outside the channel can be high. For example, in the simulations of matched guiding shown in Fig.\ \ref{fig:length_scan}, the ratios of the peak laser intensity at $2 r_\mathrm{shock}$ to that on axis is calculated to be 0.5 \%. Hence, for the guided intensities achieved in this work, the laser fields leaking beyond the shock front would be sufficient to ionize any neutral gas in this region.  Very recently, some of the present authors employed a co-axial high-order Bessel beam to increase the channel depth by ionization of neutral gas surrounding the channel formed by a lowest-order Bessel beam \cite{howard_arxiv}.

The energy transmission observed in the present work was dominated by the relatively poor coupling of the non-ideal input beam, and by the channel attenuation length of around \SI{100}{mm}. As shown in Fig.\ \ref{fig:length_scan}, an input beam matched to the lowest-order mode of the channel would increase the transmission substantially. Further, employing methods to increase the channel depth \cite{howard_arxiv} is expected to increase the attenuation length above \SI{1}{m}, leading to negligible propagation losses. Channels of this type would have an energy transmission which would equal or exceed that demonstrated by gas-filled capillary discharge waveguides \cite{Gonsalves2020}, which, for example, have demonstrated a transmission above 80\% for low-intensity pulses guided in \SI{90}{mm} long channels.

\section{Conclusions}
In summary we have demonstrated the generation of HOFI plasma channels with lengths up to \SI{100}{mm}. Optical guiding of laser pulses with a peak input intensity of $\SI{6E17}{W.cm^{-2}}$ was demonstrated over $21 z_\mathrm{R}$, in plasma channels with measured axial densities as low as $n_{e0} = \SI[separate-uncertainty=true]{1.0(3)e17}{cm^{-3}}$. The power attenuation length of this channel was measured to be $L_\mathrm{att} = \left(102\pm38\right)\SI{}{mm}$.  Guiding was also observed for lower fill pressures, for which MHD simulations predict that the axial electron density of the channels is approximately $\SI{7e16}{cm^{-3}}$. 

Measurements of the energy transmission in the channels are in good agreement with numerical simulations of beam propagation through plasma channels with transverse electron density profiles equal to those measured interferometrically. This analysis showed that the coupling efficiency, $T(0) = \left(45\,_{-11}^{+14}\right)\%$, achieved in the experiments was limited by unwanted transverse structure in the profile of the input beam. It is expected that substantially higher coupling will be possible for beams which are better matched to the lowest-order mode of the channels.

To our knowledge the plasma channels described in this work are the longest, and have the lowest on-axis density, of any free-standing plasma channel demonstrated to guide laser pulses with intensities above $>\SI{E17}{W.cm^{-2}}$. These channels were generated with only \SI{0.7}{mJ} of channel-forming laser energy per millimeter of channel, consistent with our earlier work  \cite{Shalloo:2019hv}. The ability to create long, low-density and free-standing HOFI plasma channels suggests that they are well suited to high-repetition-rate, multi-GeV plasma accelerator stages.

We would like to acknowledge the contributions of Bo Miao and Jaron Shrock to the preparation of this experiment. This work was supported by the UK Science and Technology Facilities Council (STFC UK) [grant numbers ST/P002048/1, ST/N504233/1, ST/R505006/1, ST/S505833/1]; the Engineering and Physical Sciences Research Council [studentship No.\ EP/N509711/1]; and the Central Laser Facility of the United Kingdom. L. F. and H. M. M. were supported by the U.S. Department of Energy [grant number DESC0015516] and the National Science Foundation [grant number PHY1619582]. L. C., H. J., and L. R. R. were supported by the UK Science and Technology Facilities Council (STFC UK) [grant number ST/P002056/1]. This material is based upon work supported by the Air Force Office of Scientific Research under award number FA9550-18-1-7005. This work was supported by the European Union's Horizon 2020 research and innovation programme under grant agreement No. 653782.

 \newcommand{\noop}[1]{}

\end{document}